\documentclass[runningheads]{lncse}
\usepackage{multicol} % needed for the author index
\usepackage{makeidx}  % allows for index generation
\usepackage{amsmath}
\begin{document}
\mainmatter              % start of the contributions
\addtocmark[2]{Hamiltonian Mechanics} % additional numbered TOC mark
\title{The importance of intermediate range order in silicates: molecular dynamics simulation studies}
\titlerunning{The importance of intermediate range order in silicates}
\author{J\"urgen Horbach\inst{1}, Anke Winkler\inst{1}, Walter Kob\inst{2}, and Kurt Binder\inst{1}}
\authorrunning{J. Horbach {\it et al.}}

\institute{Institut f\"ur Physik, Johannes Gutenberg--Universit\"at,\\
           D--55099 Mainz, Staudinger Weg 7, Germany
           \and
           Laboratoire des Verres, Universit\'e Montpellier II,\\
           Place E. Bataillon, cc69, 34095 Montpellier, France}

\maketitle              % typeset the title of the contribution

\begin{abstract}
We present the results of large scale computer simulations in which
we investigate the structural and dynamic properties of silicate melts
with the compositions (Na$_2$O)2(SiO$_2$) and (Al$_2$O$_3$)2(SiO$_2$).
In order to treat such systems on a time scale of several nanoseconds
and for system sizes of several thousand atoms it is necessary to use
parallel supercomputers like the CRAY T3E. We show that the silicates
under consideration exhibit additional intermediate range order as
compared to silica (SiO$_2$) where the characteristic intermediate length
scales stem from the tetrahedral network structure. For the sodium
silicate system it is demonstrated that the latter structural features
are intimately connected with a surprising dynamics in which the 
one--particle motion of the sodium ions appears on a much smaller
time scale than the correlations between different sodium ions.
\end{abstract}
\section{Introduction}
Silicate melts and glasses are an important class of materials in very
different fields, e.g.~in geosciences (since silicates are geologically
the most relevant materials) and in technology (windows, containers,
optical fibers etc.). From a physical point of view it is a very challenging
task to understand the properties of those materials on a microscopic
level, and in the last twenty years many studies on different systems
have shown that molecular dynamics computer simulations are a very
appropriate tool for this purpose \cite{angell,balucani,kob,poole}. The main
advantage of such simulations is that they give access to the whole
microscopic information in form of the particle trajectories which of
course cannot be determined in real experiments.

In {\it pure} silica (SiO$_2$) the structure is that of a disordered
tetrahedral network in which SiO$_4$ tetrahedra are connected
via the oxygens at the corners. In recent simulations we have
studied in detail various aspects of static and dynamic properties of
silica such as structural and thermodynamic properties of the glass
state~\cite{vollmayr,horbach3}, the diffusion dynamics and structural
relaxation~\cite{horbach2,horbach6,horbach1,binder}, the frequency
dependent specific heat~\cite{horbach4}, the vibrational degrees of
freedom~\cite{horbach5} and free surfaces~\cite{roder,mischler,horbach10}.
In this paper we consider silicates that contain additional
oxide components. Especially silicates with alkali oxides have
been investigated very recently in several molecular dynamics
simulations~\cite{horbach7,horbach8,horbach9,jund,banhatti,zotov,oviedo}.
We investigate here the systems (Na$_2$O)2(SiO$_2$) and
(Al$_2$O$_3$)2(SiO$_2$) (for which we use in the following the abbreviations NS2 and AS2,
respectively). Whereas sodium in NS2 plays the role of a network modifier
that partially destroys the SiO$_4$ network, aluminium in AS2 is also
a network former in that it prefers a four--fold coordination by
oxygen atoms. However, the packing of the AlO$_4$ tetrahedra is different
from that of the SiO$_4$ tetrahedra which is indicated for instance by
a different coordination number distribution of aluminium and silicon
by oxygen atoms (mainly two--fold for silicon and two-- and three--fold
for aluminium)~\cite{winkler}. As we show in the following the insertion
of sodium or aluminium atoms does not only modify the structure on local
length scales but it introduces also new intermediate length scales that
can be identified by means of the partial static structure factors.
These length scales are important for the dynamic properties as we will
demonstrate for the case of NS2.

The rest of the paper is organized as follows: In the next section we
give the main computational details and discuss the efficiency of our
simulation code on the T3E at the HLRZ Stuttgart. Then we present the
structural properties of AS2 and NS2 on intermediate length scales (Sec.~3) and
the dynamics of NS2 (Sec.~4). Eventually we summarize our results (Sec.~5).

\section{Model and details of the simulations}
In a classical molecular dynamics (MD) computer simulation one solves
numerically Newton's equations of motion for a many particle system. If
quantum mechanical effects can be neglected such simulations are able
to give in principle a realistic description of any molecular system.
The determining factor of how well the properties of a real material
are reproduced by a MD simulation is given by the potential with which
the interaction between the atoms is described. The model potential we
use to compute the interaction between the ions in NS2 and AS2 is
the one proposed by Kramer {\it et al.}~\cite{kramer} which is a
generalization of the so--called BKS potential~\cite{vanbeest} for pure
silica. It has the following functional form:
\begin{equation}
\phi_{\alpha \beta} (r)=
\frac{q_{\alpha} q_{\beta} e^2}{r} + 
A_{\alpha \beta} \exp\left(-B_{\alpha \beta}r\right) -
\frac{C_{\alpha \beta}}{r^6}\quad \alpha, \beta \in
[{\rm Si}, {\rm Al}, {\rm Na}, {\rm O}].
\label{eq1}
\end{equation}
Here $r$ is the distance between an ion of type $\alpha$ and an ion
of type $\beta$. The values of the parameters $A_{\alpha \beta},
B_{\alpha \beta}$ and $C_{\alpha \beta}$ can be found in the original
publication. The potential (\ref{eq1}) has been optimized by Kramer
{\it et al.}~for zeolites, i.e.~for systems that consist of Si,
Al, Na, O, and possible other components like phosphor. In that
paper the authors used for silicon, aluminium, and oxygen the {\it
partial} charges $q_{{\rm Si}}=2.4$, $q_{{\rm Al}}=1.9$, and $q_{{\rm
O}}=-1.2$, respectively, whereas sodium was assigned its real ion charge
$q_{{\rm Na}}=1.0$. Thus, with this set of charges charge neutrality
is fulfilled neither in NS2 nor in AS2. We have therefore modified
the Kramer potential by setting the partial charge for sodium and aluminium
to $0.6$ and $1.8$, respectively, and by introducing additional short
range potentials such that the original functional form of the Kramer
potential is approximately recovered on distances of nearest Al--O and
Na--O neighbors.  More details on the interaction potential can be found
in Refs.~\cite{horbach7,horbach8,winkler}. Our models give predictions
for structural and dynamic properties of NS2 and AS2 which are in
good agreement with experimental findings~\cite{horbach8,winkler}.
Furthermore, Ispas {\it et al.}~\cite{ispas} have shown for
(Na$_2$O)4(SiO$_2$) that {\it ab initio} simulations (Car--Parrinello
molecular dynamics) yield comparable results regarding the structure
to those obtained with molecular dynamics simulations in which our 
potential model is used.

\begin{figure}[t]
\vspace*{87mm}
\includegraphics{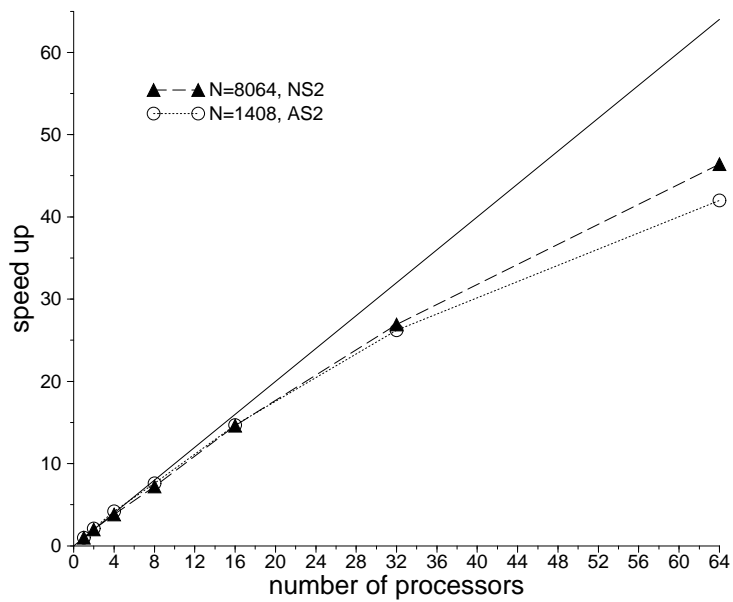}
\vspace{-0.8cm}
\caption{Speed up factor for the simulations with $N=8064$ (filled triangles)
         and $N=1408$ particles (open circles) as a function of the number of 
         processors $n$. The bisecting line (straight line) indicates a 
         perfect scaling of the performance with $n$.}
\label{fig1}
\end{figure}
The simulations have been done at constant volume: For AS2 we fixed the
density to $2.6 \, {\rm g}/{\rm cm}^3$ which is close to the experimental
density at $T=300$~K. In the case of NS2 we did simulations at the
two densities $2.37 \, {\rm g}/{\rm cm}^3$ and $2.5 \, {\rm g}/{\rm
cm}^3$, corresponding to experimental densities in the melt and at room
temperatures, respectively.  The AS2 system consists of $1480$ particles
and for the NS2 systems we used system sizes of 8064 particles at the
low density and 1152 particles at the high density.

As can be seen from Eq.~\ref{eq1} the interaction potential contains a
long--ranged Coulomb term. This part of the interaction is the most
time consuming in the calculation of the forces. To do this we made use
of the so--called Ewald summation technique~\cite{frenkel}, a method that
scales with the particle number $N$ as $N^{3/2}$. Thus, for systems
which contain about 8000 particles a huge numerical effort is required:
The longest runs (at the lowest temperatures) had a length of about
10 million time steps for which a time of two weeks was needed on
64 processors thus giving a total CPU time of about 128 weeks of (single)
processor time.

The equations of motion were integrated with the velocity form of the
Verlet algorithm. The time step of the integration was $1.6$~fs. The
temperature range investigated was $4000$~K$\ge T \ge 2100$~K in
the case of NS2 and $6100$~K$\ge T \ge 2300$~K in the case of AS2.
To equilibrate the systems the temperatures were controlled by coupling them to a
stochastic heat bath, i.e. by substituting periodically the velocities
of the particles with the ones from a Maxwell-Boltzmann distribution
with the correct temperature.  After the system was equilibrated at the
target temperature, we continued the run in the microcanonical ensemble,
i.e.~the heat bath was switched off.  We have done production runs up
to several ns real time which corresponds to several million time steps.
We have also calculated glass structures at $T=300$~K.  The glass state
was produced by cooling the system from equilibrated configurations at
our lowest temperatures with a cooling rate of about $10^{12} \, {\rm
K}/{\rm s}$. Note that we show in the following sections only results for
the lowest temperatures, i.e.~at $T=2100$~K for NS2 and at $T=2300$~K for
AS2 as well as at $T=300$~K for both systems, because the results for the
higher temperatures lead essentially to the same conclusions that we will
draw below. However, a detailed discussion of the temperature dependence
of the systems under consideration can be found in Refs.~\cite{horbach8}.

The program code was written in FORTRAN. All the parallelization was
done by using MPI subroutines. More details on the parallelization can
be found in Refs.~\cite{horbach1,winkler}. Of course, the performance
of a parallel code never scales perfectly with the number of processors
because the communication between the processors requires an additional
amount of CPU time. Fig.~\ref{fig1} shows the speed up factor on the Cray
T3E of the HLRZ Stuttgart as a function of the number of processors
$n$, i.e., the factor by which the code is faster if one uses $n$
processors instead of one.  The bisecting line indicates the limiting case where 
the communication overhead is not influenced by the speed of the code. We
see that the curves for $N=1408$ and $N=8064$ scale nearly perfectly for
$n \le 16$. For $n=64$ we obtain still a speed up factor of about $46.4$
for $N=8064$ particles whereas this factor is $42$ for $N=1408$. In most
of our simulations we have used 64 processors for the large systems and
32 processors for the small ones.

\section{Intermediate length scales in silicates}
An appropriate quantity to investigate the structure of atomic systems
on intermediate length scales is the static structure factor. It is
essentially the Fourier transform of the pair correlation function
which gives the probability of finding an atom at a distance $r$ from
another atom~\cite{balucani}. The structure factor can be directly
measured in neutron scattering experiments from the intensity of the
radiation observed with a momentum transfer $\hbar {\bf q}$ ($\hbar$:
Planck's constant, ${\bf q}$: wave--vector of the momentum transfer). In
a three--component system one can define six partial structure factors
as~\cite{balucani}
\begin{equation}
   S^{\alpha \beta}(q) = 
     \frac{1}{N}
     \sum_{k=1}^{N_{\alpha}} \sum_{l=1}^{N_{\beta}}
     \left< \exp \left( i {\bf q} \cdot [ {\bf r}_k - {\bf r}_l ] \right)
     \right> .  \label{eq2}
\end{equation}
where the first sum runs over $N_{\alpha}$ particles of type $\alpha$ and
the second one over $N_{\beta}$ particles of type $\beta$.

\begin{figure}[t]
\vspace*{87mm}
\includegraphics{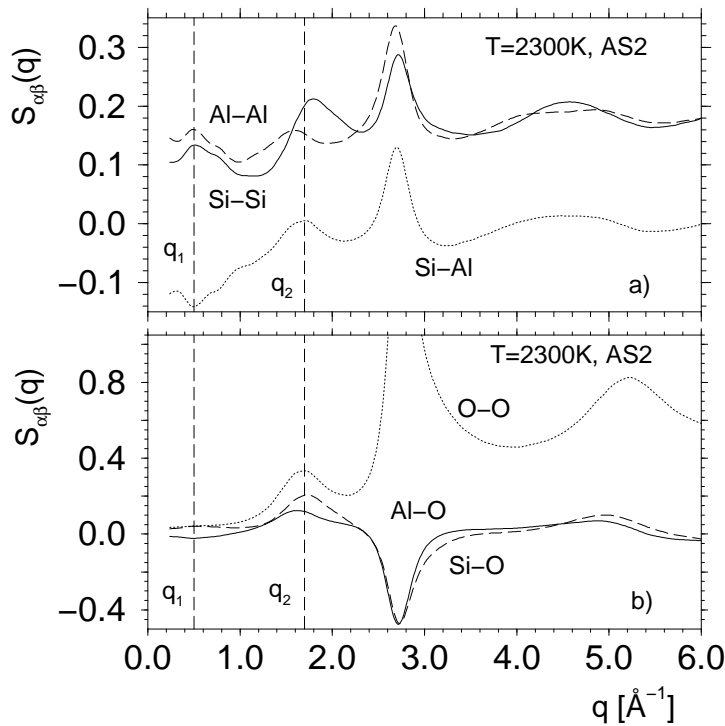}
\vspace*{0.5cm}
\caption{Partial static structure factors for AS2 at $T=2300$~K. 
         a) Al--Al, Si--Si, and Si--Al correlations, b) Si--O, 
         Al--O, and O--O correlations. For the meaning of the dashed
         vertical lines see text.
         \label{fig2}}
\end{figure}
\begin{figure}[htp]
\vspace*{-0.2cm}
\includegraphics{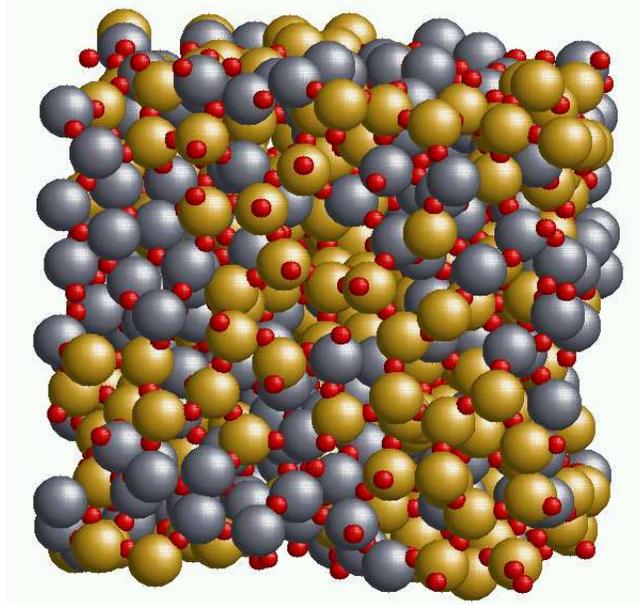}
\vspace*{7.7cm}
\caption{Snapshot of (Al$_2$O$_3$)2(SiO$_2$) (AS2) at $T=300$~K. The 
         size of the spheres is chosen
         such that one can identify aluminium-- and silicon--rich regions:
         The aluminium and oxygen atoms are shown respectively as big blue 
         and gold spheres, whereas the oxygen atoms are shown as small red spheres.
         \label{fig2a}}
\end{figure}
Fig.~\ref{fig2} shows $S_{\alpha \beta}(q)$ for AS2 at the temperature
$T=2300$~K. For $q>2.3$~\AA$^{-1}$ the partial structure factors reflect
length scales of nearest neighbors. In AS2 the smallest distances between
atoms are those of Al--O and Si--O bonds that have lengths of about $1.6$
to $1.65$~\AA.  The peaks around $q_2=1.7$~\AA$^{-1}$ in $S_{\alpha
\beta}(q)$ (marked by dashed vertical lines in Fig.~\ref{fig2}) are due
to the order that arises from the tetrahedral network structure. The
length scale $2\pi/1.7$~\AA$^{-1}=3.7$~\AA~that corresponds to
this peak is approximately the spatial extent of connected SiO$_4$ and
AlO$_4$ tetrahedra.  Note that in silica a peak at $1.7$~\AA$^{-1}$
is also a very prominent feature and is called there first sharp
diffraction peak.  But in contrast to silica one observes in AS2 an
additional peak at $q_1=0.5$~\AA$^{-1}$ in the Al--Al, Si--Si, and
Si--Al correlations and only weakly pronounced also in the remaining
correlations in which oxygen is involved. $q_1$ corresponds to a length
scale of about $12.5$~\AA~and has its reason in a slightly different
ordering of AlO$_4$ complexes as compared to the SiO$_4$ network (for
details see \cite{winkler}).  This relatively large length scale shows
that large system sizes are required to analyze the structure of systems
like AS2 in a sensible way.  The different ordering of AlO$_4$ leads to
a structure where an AlO$_4$ tetrahedron prefers to be surrounded on a
local scale by other AlO$_4$ tetrahedra. This leads to a structure where
AlO$_4$ complexes are connected to each other as string--like objects
through the system that form a percolating network.  This is illustrated
by the snapshhot in Fig.~\ref{fig2a} where the aluminium and silicon
atoms are shown as the blue and gold spheres, respectively. Note that it
does not matter that this snapshot is at $T=300$~K and not at $T=2300$~K
as the structure factors in Fig.~\ref{fig2} because we find only small
differences in structural quantities at both temperatures.  Thus, we see
that the aluminium atoms are not at all homogeneously distributed and
if one only considers the Al atoms voids with a size of about $2 \pi /
q_1$ are found that lead to the peak at $q_1$ in $S_{{\rm Al-Al}}$. It
is not surprising that these voids are also reflected in the Si--Si and
Si--Al correlations but much less in the correlations containing oxygen
(as can be seen in Fig.~\ref{fig2}b): The oxygen atoms are essentially
homogeneously distributed on the length scale $2 \pi / q_1$ since they
are nearest neighbors of silicon and aluminium with a similar
length of Si--O and Al--O bonds.

\begin{figure}[t]
\vspace*{87mm}
\includegraphics{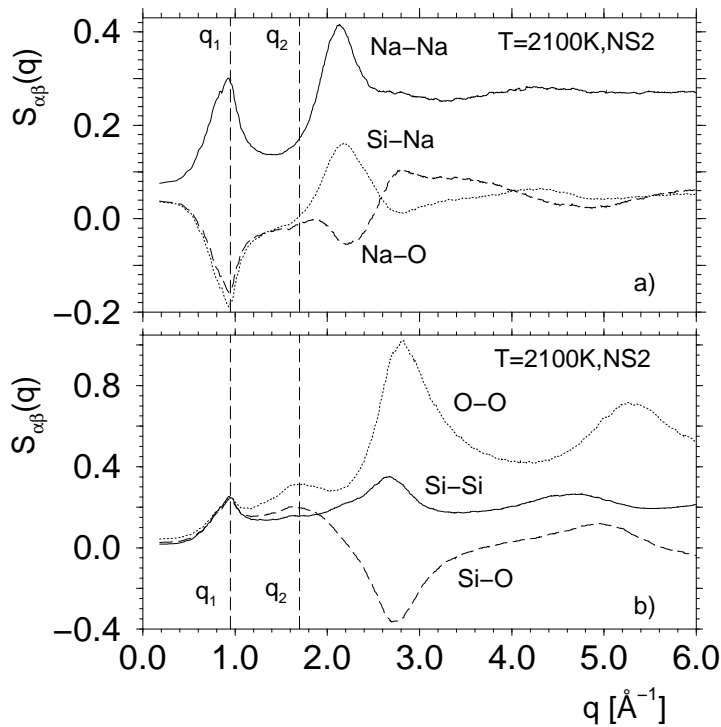}
\vspace*{0.5cm}
\caption{Partial static structure factors for NS2 at $T=2100$~K. 
         a) Na--Na, Si--Na, and Na--O correlations, b) Si--O, 
         Si--Si, and O--O correlations. For the meaning of the dashed
         vertical lines see text.
         \label{fig3}}
\end{figure}
The sodium ions in NS2 play a different role from the aluminium atoms in
AS2 since they partially destroy the SiO$_4$ network. This can be directly
recognized in the partial structure factors for NS2 which are shown in
Fig.~\ref{fig3} at $T=2100$~K for the density $\rho=2.37$~g$/$cm$^3$: The
peak at $q_2=1.7$~\AA$^{-1}$ that reflects the structure of a tetrahedral
network is absent in the correlations with sodium (Fig.~\ref{fig3}a) and
is especially in $S_{{\rm Si-Si}}$ much weaker pronounced than in AS2
(Fig.~\ref{fig3}b).  But we observe again a second prepeak at smaller
$q$, now around $q_1=0.95$~\AA$^{-1}$. This $q$ value is of the order of the
length scale of next nearest Na--Na or Si--Na neighbors (around $6.6$~\AA).
Again, the peak at $q_1$ is the characteristic wave--vector of a 
percolating network that is now formed by the sodium atoms. At first glance
it seems to be surprising that also $S_{{\rm O-O}}$ exhibits a peak
at $q_1$. But the role of oxygens is different in NS2 from that in AS2:
The nearest neighbor distance for Na--O, $2.2$~\AA, is larger than for 
Si--O which is $1.6$~\AA. And the arrangement of oxygen around sodium
is different from the tetrahedral one around silicon (for more details
see Ref.~\cite{horbach8}). Thus, the distribution of oxygen atoms in NS2
is not homogeneous on the length scale $2 \pi / q_1$.

\begin{figure}[t]
\vspace*{87mm}
\includegraphics{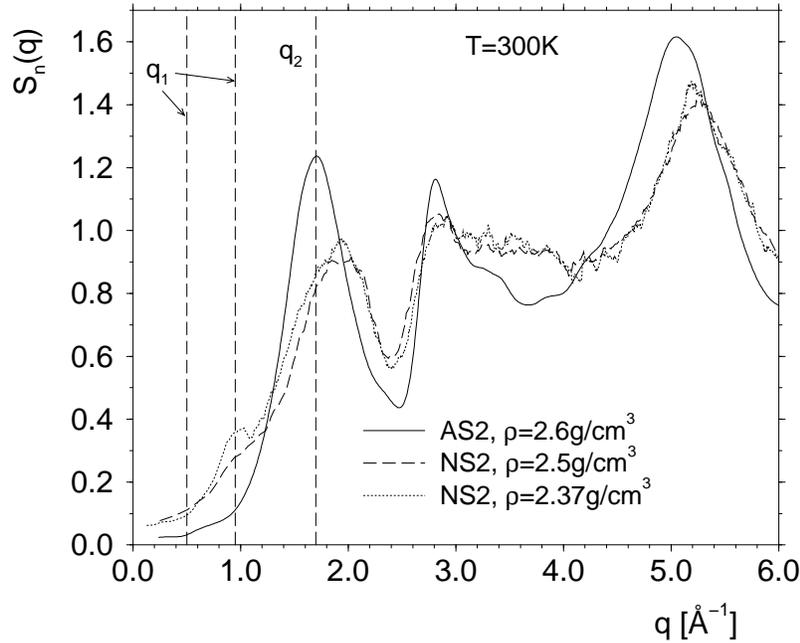}
\caption{$S_{{\rm n}}(q)$ at $T=300$~K for AS2 and for NS2 at the indicated densities.
         The dashed vertical lines mark the position of the peaks at 
         $q_1=0.5$~\AA$^{-1}$ in AS2, $q_1=0.95$~\AA$^{-1}$ in NS2,
         and $q_2=1.7$~\AA$^{-1}$ in both systems. 
         \label{fig4}}
\end{figure}
So far we have seen that NS2 and AS2 exhibit intermediate order on a
relatively large length scales. This gives rise to a prepeak in
$S_{\alpha \beta}(q)$ at $q_1$ which is $0.5$~\AA$^{-1}$ for AS2 and
$0.95$~\AA$^{-1}$ for NS2. But does one see these peaks at $q_1$ also 
in experiments?
In experiments such as neutron scattering one does not have access to
the partial structure factors for systems like NS2 or AS2. Here one
measures a sum of the partial structure factors whereby the different
contributions are weighted by the neutron scattering lengths $b_{\alpha}$:  
\begin{equation}
   S_{{\rm n}}(q) = 
     \frac{1}{\sum_{\alpha} N_{\alpha} b_{\alpha}^2}
     \sum_{kl} b_k b_l 
     \left< \exp \left(i {\bf q} \cdot [ {\bf r}_k - {\bf r}_l ] \right)
     \right> .  \label{eq3}
\end{equation}
The values for $b_{\alpha}$ are $0.4149 \cdot 10^{-12}$~cm, $0.3449 \cdot
10^{-12}$~cm, $0.363 \cdot 10^{-12}$~cm, and $0.5803 \cdot 10^{-12}$~cm
for silicon, aluminium, sodium, and oxygen, respectively~\cite{sears}.
By weighting the $S_{\alpha \beta}(q)$ from our simulation with the
$b_{\alpha}$ in accordance with Eq.~(\ref{eq3}) one can easily calculate
the quantity $S_{{\rm n}}(q)$.  It is shown in Fig.~\ref{fig4} at
$T=300$~K for NS2 at the two densities $\rho= 2.37$~g$/$cm$^3$ and
$2.5$~g$/$cm$^3$ and for AS2 at $\rho= 2.6$~g$/$cm$^3$. We infer from
this figure that the aforementioned prepeaks at $q_1$ can be seen
in AS2 and in NS2 at the higher density only as a weakly pronounced
shoulder. Thus it would be difficult to identify them in a neutron
scattering experiment. Only in NS2 at the lower density one can clearly
see the prepeak at $q=0.95$~\AA$^{-1}$. But at this density we observe
a negative pressure of about $-1.6$~GPa at $T=300$~K, a condition that would be difficult
to realise in an experiment. However, in an experiment under normal
pressure conditions the density decreases if one
goes to higher temperatures. And indeed, very recent neutron scattering
experiments of Meyer {\it et al.}~do find the feature at $q_1$ 
for NS2~\cite{meyer}. Meyer {\it et al.}~have measured for the first time
the temperature dependence of the structure factor from $T=300$~K (where the
system is in a glass state) to $T=1600$~K (where one has a melt). 
They find that the feature at $q_1$ becomes more and more pronounced by
increasing the temperature and one can clearly identify it at $T=1600$~K.
This behavior is similar to what we see in our simulations and can be
understood by an decreasing density in the experiment if the temperature
is increased.

\section{The dynamics of NS2}
In a recent simulation we have demonstrated that the dynamics in
in NS2 is much faster than the one in pure silica~\cite{horbach8}. Even at a relatively high
temperature of $T=2750$~K the diffusion constants of silicon and oxygen
are two orders of magnitude larger in NS2 than in SiO$_2$. Furthermore, in NS2
the sodium diffusion decouples more and more from the silicon and oxygen
diffusion such that at temperatures $T\le 2500$~K the dynamics of the Na
atoms is about two orders of magnitude faster than the one of the
oxygen and silicon atoms~\cite{horbach8}. This is in qualitative agreement
with the expermental fact that NS2 is an ion conducting material.

\begin{figure}[t]
\vspace*{87mm}
\includegraphics{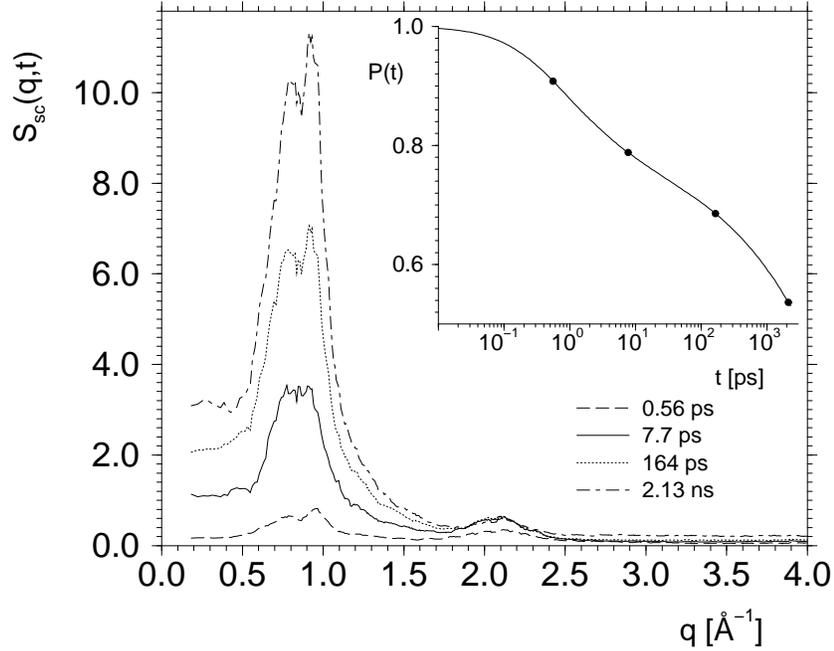}
\vspace{0.5cm}
\caption{Swiss cheese structure factor $S_{{\rm sc}}(q,t)$ at the indicated times. The inset
         shows the probability $P(t)$ (see text). The circles on the
         curve for $P(t)$ are at the times at which $S_{{\rm sc}}(q,t)$
         is shown.
         \label{fig5}}
\end{figure}
Thus, since essentially the Si and O atoms do not move with respect to the 
movement of the Na atoms one may expect that sodium diffusion is restricted
to a small subspace of the configuration space. The Si and O atoms form
a quasi--frozen matrix for the Na atoms and it would be surprizing if
the sodium atoms are able to diffuse into this matrix. In order to check
this idea we have calculated a (coarse grained) probability of finding
{\it no} sodium atom at a given location in space. Following the approach of Jund
{\it et al.}~\cite{jund} we calculate this probability by dividing the 
system into $48^3$ cubes (of length $L/48 \approx 1.01$~\AA). Then we calculate
the probability $P(t)$ that a cube which does not contain a sodium ion
at time zero is also not visited by a sodium ion until a later time $t$.
The time dependence of $P(t)$ is shown in the inset of Fig.~\ref{fig5}. 
From this graph we recognize that after 2.5~ns, i.e.~after more than the
$\alpha-$relaxation time of the matrix~\cite{horbach8}, more than 50\%
of the cubes have not yet been visited by a sodium atom. (We mention
that after this time the mean squared displacement of the Na atoms is
more than (45~\AA)$^2$, which shows that these atoms have moved a large
distance. On this time scale also the {\it local} structure of the Si--O matrix
is partially reconstructed~\cite{horbach8}.)  Hence we can conclude that on this time
scale the sodium free region forms a percolating structure
around a network of channels, i.e.~it has somewhat the structure of a
Swiss cheese.  In order to investigate the structure of this percolating
region we define a ``Swiss cheese'' structure factor $S_{\rm sc}(q,t)$
as follows: We assign to each cube which has not been visited by
a sodium atom until time $t$ a point and we compute the static structure
factor of $N_{{\rm sc}}(t)=P(t) (48^3-N_{{\rm Na}})$ points:
\begin{equation}
   S_{\rm sc}(q,t)= \frac{1}{N_{{\rm sc}}(t)} \sum_{k,l=1}^{N_{{\rm sc}}(t)}
      \left< \exp(i \vec{q} \cdot (\vec{r}_k - \vec{r}_l))  \right> \ .
\end{equation}
\label{eq4}
This quantity is shown in Fig.~\ref{fig5} for four different times:
$t=0.56$~ps, $7.7$~ps, $164$~ps, and $2.13$~ns.  We see that $S_{\rm
sc}(q,t)$ has a pronounced peak at $q_1=0.95$~\AA$^{-1}$
which is also a prominent feature in $S_{\rm Na-Na}(q)$, as we have
seen in the preceding section. Hence we can now conclude that the peak at
$q_1$ in $S_{\rm Na-Na}(q)$ corresponds to the typical distance between
the channels. Note that with increasing time the height of this peak
increases quickly. However, it is clear that the peak at $q_1$ decreases
again on the time scale on which the matrix starts to reconstruct itself
significantly and thus rearranges the channel structure.

\begin{figure}[t]
\vspace*{87mm}
\includegraphics{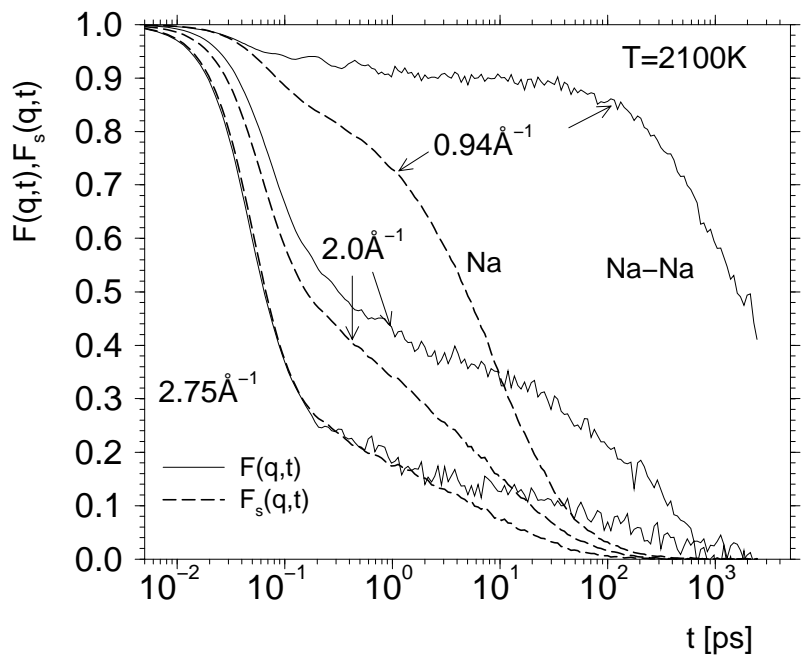}
\vspace*{-0.2cm}
\caption{Coherent intermediate scattering functions $F(q,t)$ for sodium--sodium correlations
         (bold solid lines) and incoherent intermediate scattering functions 
         $F_{{\rm s}}(q,t)$ (dashed lines) at $T=2100$~K for the indicated values
         of $q$.
         \label{fig6}}
\end{figure}
We address now the question how the sodium ions relax inside the
channels. An appropriate quantity to investigate this point are
time dependent density--density correlation functions, i.e.~the
coherent intermediate scattering function $F(q,t)$ and its self
part, the incoherent intermediate scattering function $F_{{\rm
s}}(q,t)$~\cite{balucani}.  In Fig.~\ref{fig6} we show $F(q,t)$ for
the Na--Na correlations (solid lines) as well as $F_{\rm s}(q,t)$
for the sodium atoms (dashed lines) for three different wave--vectors:
$q=0.94$~\AA$^{-1}$, $2.0$~\AA$^{-1}$, and $2.75$~\AA$^{-1}$. From this
figure we infer immediately a surprising result: At $q=0.94$~\AA$^{-1}$,
i.e.~at the characteristic $q$ value of the sodium channel structure,
$F(q,t)$ decays on a time scale which is two orders of magnitude larger
than the one for $F_{\rm s}(q,t)$. Such a strong difference cannot be
explained by a de Gennes narrowing argument~\cite{balucani}. Instead
this result can be rationalized by the idea that the sodium atoms move
quickly between preferential sites, since this type of motion gives rise
to a fast decorrelation of the incoherent function whereas it does not
affect the coherent one. Only on the time scale of the relaxation of the
SiO$_2$ matrix also the coherent function starts to decay.  Note that
the slow decay of $F(q,t)$ is found only for wave--vectors around
0.95~\AA$^{-1}$. For different $q$ the function decays significantly
faster as can be seen from the other curves shown in Fig.~\ref{fig6}.

More details on the issues discussed in this section can be found in 
Ref.~\cite{horbach9}.

\section{Summary}
Large scale molecular dynamics computer simulations as the ones presented in this
paper for AS2 and NS2 require the use of parallel computers such as the Cray T3E.
Only then it is possible to simulate these systems on a scale of several ns for
system sizes which are big enough to study the structure and dynamics on
intermediate length scales (up to 8000 particles in our case). Although no 
neutron scattering experiments can be done yet
for temperatures above 2000~K, the structural and dynamical properties are already
present at these high temperatures and thus, one can gain insight into features
that one observes in experiments. Furthermore, this insight is much more detailed
in a MD simulation than in an experiment since one has access to the full microscopic
information in form of the particle trajectories.

We have exploited this fact for the case of AS2 and NS2 by showing that these systems exhibit
intermediate range order on length scales that are larger than the one given
from the tetrahedral network structure in {\it pure} silica. The reason for this
is a different ordering of Al--O and Na--O complexes and leads to a percolating
network of these structural elements through the SiO$_4$ network. We have shown
for the example of NS2 that this intermediate range order is also important to understand
the dynamics: In NS2 the sodium ions that move through channels in the Si--O matrix
and the structure of these channels is connected with the prepeak in the static
structure factor at $0.95$~\AA$^{-1}$. The presence of these channels leads to
a surprising decoupling of the fast (single particle) sodium motion from correlations between
different sodium atoms that decay on the time scale of the channel relaxation. 

Acknowledgments:
We thank the HLRZ Stuttgart for a generous grant of computer time 
on the CRAY T3E. A. W. is grateful to SCHOTT Glas for partial
financial support.

\clearpage
\addtocmark[2]{Author Index}% additional numbered TOC entry
\markboth{Author Index}{Author Index}%
\renewcommand{\indexname}{Author Index}%
\threecolindex  % starts the next index in three column mode
\printindex
\clearpage
\addtocmark[2]{Subject Index} % additional numbered TOC entry
\markboth{Subject Index}{Subject Index}
\renewcommand{\indexname}{Subject Index}
\makeatletter
%\@input@{subjidx.ind}
\end{document}